\documentclass[twocolumn,amsmath,amssymb,aps,groupedaddress,nofootinbib]{revtex4-1}
\usepackage{amsthm}

\newcommand{\ket}[1]{|{#1}\rangle}

\begin{document}

\title{Quantum fault tolerance in small experiments}

\author{Daniel Gottesman}
\email{dgottesman@perimeterinstitute.ca}
\affiliation{Perimeter Institute, Waterloo, Canada;
    CIFAR QIS Program, Toronto, Canada}

\begin{abstract}
I discuss a variety of issues relating to near-future experiments demonstrating fault-tolerant quantum computation.  I describe a family of fault-tolerant quantum circuits that can be performed with 5 qubits arranged on a ring with nearest-neighbor interactions.  I also present a criterion whereby we can say that an experiment has succeeded in demonstrating fault tolerance.  Finally, I discuss the possibility of using future fault-tolerant experiments to answer important questions about the interaction of fault-tolerant protocols with real experimental errors.
\end{abstract}

\maketitle

\section{Introduction}

Experimental control of small quantum computers has reached superb levels of accuracy, with two-qubit gate errors in some cases below $0.1\%$~\cite{Wineland}.  However, large quantum algorithms can easily involve $10^9$ or more gates, so it is extremely likely that an error will occur sometime during the algorithm, even if there are significant further increases in accuracy.  To successfully complete a large quantum computation will therefore require the computation to be encoded in a fault-tolerant protocol, so that the final answer extracted from the computation is correct despite errors occurring at a low rate during the computation.

While there have been demonstrations of quantum error-correcting codes in a number of different experimental systems (e.g.,~\cite{KnillQECC,PanQECC,LaflammeQECC,BlattQECC,RarityQECC,IBMQECC}), a full fault-tolerant protocol has not yet been demonstrated.  In quantum error correction experiments to date, the state is encoded, left to wait some amount of time, then decoded and corrected.  If there is a single error during the waiting stage (and sometimes the experimentalists will introduce errors artificially during this time), the code can correct that error.  However, the encoding, decoding, and correction are done with quantum circuits that lack the extra protection of a fault-tolerant circuit.  If a single error occurs during any of these steps, the decoded state will potentially be erroneous; in contrast, a fault-tolerant circuit would give the correct output no matter when the error occurs.%
\footnote{The simplest fault-tolerant protocols will still fail if there are two errors in the course of the circuit.  More sophisticated protocols are possible which have a threshold error rate~\cite{AB97,Kit97,KLZ98}, and the circuit is protected provided the error rate per gate or time step is below the threshold value.}
In addition, fault-tolerant protocols offer the possibility of safely performing gates on the encoded states rather than just storing quantum information.

Since fault tolerance is likely to be an essential part of a large quantum computer, demonstrating fault tolerance is an important benchmark for any implementation of quantum computation.  Given the levels of experimental control that have been reached, it is quite plausible that experiments meeting this benchmark could be done in the near future.  It therefore behooves us to examine what would qualify as an experimental realization of fault tolerance, how it can be done, and what the difficulties are.  In this letter, I discuss these issues and present the smallest circuits that I would consider to be a demonstration of fault tolerance, which involve $5$ qubits arranged in a ring.

\section{Criterion for experimental fault tolerance}

The first question to consider is ``When can an experiment on a small system be said to have demonstrated fault tolerance?''  My answer is as follows: Compare the error rate for an unencoded circuit to that of an encoded version of the same circuit on the same hardware.  For a successful demonstration of fault tolerance, the encoded circuit must have a lower error rate for all circuits in the family of circuits of interest.  While the answer seems straightforward, there are many subtle points in this definition, and additional practical complications with realizing this criterion.  I will discuss these below without giving a full resolution for some of the more difficult points.  Nevertheless, for small systems, it should be possible to make a convincing demonstration of fault tolerance.

For the current purposes, it is best to restrict attention to \emph{complete quantum circuits}, circuits that start by preparing qubits in a standard state (e.g., $\ket{0}$), perform a sequence of gates on those qubits, and then terminate by measuring one or more qubits to get classical output bits.  I will also allow circuits that prepare qubits later on, once the circuit has already started, and circuits that measure somewhere in the middle of the circuit, and perhaps adapt the future gates in the circuit depending on the measurement results.  It is important to consider complete circuits because ultimately, the experiments that are being done have a classical input (instructions provided by the experimenter) and a classical output (the data gathered at the end of the experiment).  The input here becomes subsumed into the choice of circuit to perform, so we can think of complete circuits as ones with no inputs and that produce a classical probability distribution as output.  Since the initialization of the experiment and the final measurements used to collect the data are themselves susceptible to error, leaving them out of our fault-tolerant protocol can lead to a highly deceptive impression of the robustness of the system against errors.

A \emph{circuit encoding} is a mapping from some family of complete quantum circuits (the \emph{original circuit}) to complete quantum circuits (the \emph{encoded circuit}), along with a rule for interpreting (\emph{decoding}) the classical output of the encoded circuit.%
\footnote{Normally, we would require that the encoding map and the decoding procedure be efficiently computable classically.  Since we are dealing with only small circuits here, ``efficient'' is not actually well-defined, being an asymptotic property, so we simply apply a rough condition that the encoding and decoding not be too complicated.  In most cases, the encoding map simply replaces each element (state preparation, gate, or measurement) of the original circuit with a \emph{gadget} implementing that element on a codeword of a quantum error-correcting code, perhaps adding some additional error correction steps.  Another subtlety is that for small circuits, the correct output distribution can be computed on existing classical computers, so we should not allow the circuit encoding to replace part or all of the quantum circuit with classical processing.}
We require that if the original circuit and the encoded circuit are both implemented in an ideal manner --- with no possibility of errors --- then the decoded output distribution for the encoded circuit must be the same as the output distribution for the original circuit.%
\footnote{Here, I require that the distributions be exactly the same.  One could allow them to simply be very close together, but that is an unnecessary complication here, as most fault-tolerant protocols allow an exact realization, provided the gates used in the original circuit come from an appropriate set of fault-tolerant friendly gates.}
Note that the circuits might not have a deterministic outcome, so we require that the probability distribution of outcomes are the same.  This includes matching correlations between output bits in case multiple qubits are measured.

We don't require that a circuit encoding work for all possible original circuits, only circuits drawn from some family.  In the usual literature of fault tolerance, we are concerned with \emph{universal} families, i.e., families of circuits capable of implementing arbitrary quantum algorithms.  For the purposes of demonstrating fault tolerance, it is also interesting to consider smaller families of circuits which do not involve a universal set of gates, including some very simple families of circuits.  For reasons to be discussed shortly, we need to restrict attention to families where the encoded circuits involve a fixed number of qubits, and this also restricts the ability of the circuits to perform arbitrary algorithms.  The question of how to choose a suitable family is a subtle one (and maybe one that does not have an objective answer at all), but the family of circuits should contain a variety of different types of circuit elements, both small and large circuits, and be large enough to provide a variety of different circuits.  Otherwise, we have no assurance that we are testing the full protocol for fault tolerance and not just some specific applications of it.

To determine if a circuit encoding is fault tolerant, we should, ideally, then go through every circuit $C$ in the family and implement both the original version of $C$ and the encoded version of $C$ in the experimental system.  We then need to compare the error rates for these two versions of $C$.  The error rate of the original circuit is the statistical distance $\frac{1}{2} \sum_i |p_i-q_i|$ between the output distribution $\{q_i\}$ of the original circuit implemented in a real system, with errors, and the output distribution $\{p_i\}$ of the original circuit if it could be implemented in an ideal way.  The error rate of the encoded circuit is the distance between the ideal output distribution of the original circuit and the decoded output distribution of the actual implementation of the encoded circuit.  Any reasonable distance measure between classical probability distributions would work instead of the statistical distance.

It is only really meaningful to compare error rates between circuits implemented in the same system.  Otherwise, any effect of the circuitry can be easily outweighed by differing error rates between the systems.  Since the error rates could differ on different physical qubits, it is important to implement the original circuit using the best qubit(s) available in the system.  The encoded circuit can be implemented in whichever way seems suitable.  The goal is to show that the encoding improves the error rate over any approach that does not encode the qubits, not that it improves over some particularly poor way of performing the original circuit.

Note that to achieve fault tolerance, the circuit encoding must reduce the error rate for \emph{all} circuits in the family.  A circuit encoding that improves the error rates for some circuits in the family but not for others is not demonstrating a real improvement.  For instance, previous experiments on quantum error correction can be viewed in this framework as working on the family of circuits which prepare a qubit, wait varying amounts of time, and then measure.  However, these experiments only produce a lower error rate for long time periods, where the dominant errors come from the waiting time, and not for short ones, where errors in preparation and measurement are important.  The preparation and measurement steps are not done in a fault-tolerant way, and this is revealed by considering short circuits as well as long ones.  This example also emphasizes the importance of picking an appropriate family of circuits to study, since varying over circuits in the family can help to emphasize different circuit elements.

\section{Small circuits for fault tolerance}

In order to perform a demonstration of fault tolerance with the smallest possible circuits, we are going to need a small quantum error-correcting code that also has a small fault-tolerant protocol. A number of codes are plausible candidates, including the $5$-qubit code~\cite{fivequbit1,fivequbit2}, the $7$-qubit code~\cite{CalderbankShor,Steane}, surface codes~\cite{Kitaev,DKLP}, and the $9$-qubit Bacon-Shor subsystem code~\cite{ninequbit,Bacon,Poulin}, but all of these would require $10$ or more qubits for a full fault-tolerant protocol.  The Supplementary Material~\cite{Supplementary} has a longer discussion of the advantages and disadvantages of these codes for fault tolerant experiments.

Luckily, there is a better option available.  The $4$-qubit code~\cite{four-qubit1,four-qubit2} is a quantum error-detecting code: It can detect any single-qubit error, but cannot be used to correct the error that is detected.  It is a small code with a simple fault-tolerant protocol~\cite{ShorFT}.  However, because it cannot correct an error, it can only improve error rates via post-selection.  Demonstrations that rely on post-selection can be controversial, but the use of post-selection is actually common for ancilla preparation as a subroutine in large fault-tolerant protocols.  Thus, at a minimum, a successful demonstration of fault tolerance with the $4$-qubit code could be considered as a demonstration of fault-tolerant ancilla preparation.

\begin{figure}
\begin{picture}(60,50)
\put(18,46){\circle*{4}}
\put(42,46){\circle*{4}}
\put(11,24){\circle*{4}}
\put(49,24){\circle*{4}}
\put(30,10){\circle*{4}}

\put(18,46){\line(1,0){24}}
\put(11,24){\line(1,3){7}}
\put(49,24){\line(-1,3){7}}
\put(11,24){\line(3,-2){19}}
\put(49,24){\line(-3,-2){19}}

\put(30,0){\mbox{A}}
\put(1,20){\mbox{$1$}}
\put(8,46){\mbox{$2$}}
\put(47,46){\mbox{$3$}}
\put(54,20){\mbox{$4$}}
\end{picture}
\caption{$5$ qubits arranged in a ring, with nearest-neighbor interactions indicated.  Qubits $1$ through $4$ are numbered and the ancilla qubit is marked ``A''.}
\label{fig:ring}
\end{figure}
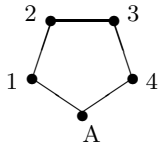

We will use $5$ qubits arranged in a ring, as in figure~\ref{fig:ring}.
The four-qubit code encodes two logical qubits, with the basis codewords
\begin{align}
\ket{00} \rightarrow& \ket{0000} + \ket{1111} \\
\ket{01} \rightarrow& \ket{1100} + \ket{0011} \\
\ket{10} \rightarrow& \ket{1010} + \ket{0101} \\
\ket{11} \rightarrow& \ket{0110} + \ket{1001}.
\end{align}
It has stabilizer generated by $X \otimes X \otimes X \otimes X$ and $Z \otimes Z \otimes Z \otimes Z$.  

There are a number of gates that can be performed fault-tolerantly on this code without using any extra qubits.  The logical Pauli gates on the two encoded qubits are just tensor products of single-qubit Paulis: $X$ on the first qubit is $X \otimes I \otimes X \otimes I$ and $X$ on the second logical qubit is $X \otimes X \otimes I \otimes I$, while $Z$ on the first logical qubit is $Z \otimes Z \otimes I \otimes I$ and $Z$ on the second logical qubit is $Z \otimes I \otimes Z \otimes I$.  Performing the Hadamard transform on all four physical qubits $H \otimes H \otimes H \otimes H$ does the Hadamard transform on both logical qubits but also switches them (so the first logical qubit becomes the second and vice-versa).  It is even possible to do a two-qubit gate on this code.  Let $R = \mathrm{diag}(1,i)$ be the $\pi/4$ phase rotation.  Then $R \otimes R \otimes R \otimes R$ does the logical controlled-$Z$ gate $\mathrm{diag}(1,1,1-1)$ between the two logical qubits, followed by $Z \otimes Z$ on the logical qubits.  We can therefore consider families of circuits using any of these logical gates.

Measuring the two logical qubits in the standard basis is straightforward: Simply measure all four qubits.  If the output bits have an odd number of $1$s, an error is detected, and we should discard the run.  Otherwise, the output string is one of the eight strings $0000$, $1111$, $1100$, $0011$, $1010$, $0101$, $0110$, or $1001$, and we decode it by seeing which of the four logical basis states that string appears in.

The most difficult part of the fault-tolerant protocol is preparing the initial state.  There are two initial states that can be directly prepared fault-tolerantly.  The more straightfoward one is the logical $\ket{0+}$ state, which is just $(\ket{00} + \ket{11}) \otimes (\ket{00} + \ket{11})$, two Bell states, each of which can be prepared from $\ket{00}$ with a Hadamard and a CNOT.  This encoding circuit is fault tolerant: A gate error during a CNOT could mess up one of the two Bell pairs, but an error on one qubit of a maximally entangled state is equivalent to an error on the other qubit.  Thus, a single CNOT error results in a state which is only wrong on one qubit --- a detectable error.  This initial state preparation fulfills the conditions of fault tolerance, and is actually even simpler than our goal, using only $4$ qubits in total, but it is a bit too trivial and doesn't truly test the performance of a fault-tolerant protocol in the experimental system.  Almost as trivial is the logical Bell state $\ket{00} + \ket{11}$, which is also two physical Bell states between qubits $2$ and $3$ and qubits $1$ and $4$.  (The latter can be prepared as a Bell state between qubits $1$ and $A$, with $A$ then swapped with $4$.)

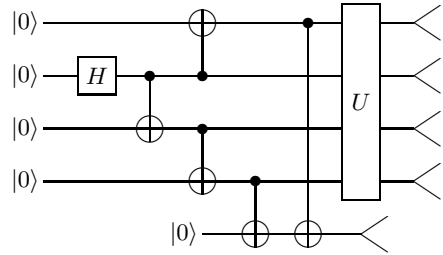
\begin{figure}
\begin{picture}(200,120)
\put(20,100){\line(1,0){113}}
\put(20,80){\line(1,0){13}}
\put(33,73){\framebox(14,14){$H$}}
\put(47,80){\line(1,0){86}}
\put(20,60){\line(1,0){113}}
\put(20,40){\line(1,0){113}}

\put(60,80){\circle*{4}}
\put(60,80){\line(0,-1){25}}
\put(60,60){\circle{10}}

\put(80,80){\circle*{4}}
\put(80,80){\line(0,1){25}}
\put(80,100){\circle{10}}

\put(80,60){\circle*{4}}
\put(80,60){\line(0,-1){25}}
\put(80,40){\circle{10}}

\put(80,20){\line(1,0){60}}
\put(100,40){\circle*{4}}
\put(100,40){\line(0,-1){25}}
\put(100,20){\circle{10}}
\put(120,100){\circle*{4}}
\put(120,100){\line(0,-1){85}}
\put(120,20){\circle{10}}
\put(140,20){\line(3,2){10}}
\put(140,20){\line(3,-2){10}}

\put(133,33){\framebox(14,74){$U$}}
\put(147,100){\line(1,0){13}}
\put(147,80){\line(1,0){13}}
\put(147,60){\line(1,0){13}}
\put(147,40){\line(1,0){13}}

\put(160,100){\line(3,2){10}}
\put(160,100){\line(3,-2){10}}
\put(160,80){\line(3,2){10}}
\put(160,80){\line(3,-2){10}}
\put(160,60){\line(3,2){10}}
\put(160,60){\line(3,-2){10}}
\put(160,40){\line(3,2){10}}
\put(160,40){\line(3,-2){10}}

\put(6,93){\makebox(14,14){$\ket{0}$}}
\put(6,73){\makebox(14,14){$\ket{0}$}}
\put(6,53){\makebox(14,14){$\ket{0}$}}
\put(6,33){\makebox(14,14){$\ket{0}$}}
\put(66,13){\makebox(14,14){$\ket{0}$}}

\end{picture}
\caption{Fault-tolerant protocol for the $4$-qubit code.  $U$ can be any sequence of the fault-tolerant gates listed in the text.  The output is discarded if the last (A) qubit gives the measurement result $1$ or if the measurement results for the other four qubits give an odd number of $1$s.  Otherwise the two logical output bits are given by the parity of the first two output bits and the first and third output bits.}
\label{fig:fourFT}
\end{figure}

A more interesting initial state to prepare is the logical $\ket{00}$ state.  The ``cat'' state $\ket{0000} + \ket{1111}$ can be prepared from four physical $\ket{0}$ states using a Hadamard transform and three CNOTs, for instance as shown in figure~\ref{fig:fourFT}.  However, without additional checking, this circuit is not fault-tolerant: An error during a CNOT could cause two errors in the state, for instance resulting in $\ket{1100} + \ket{0011}$, the logical $\ket{01}$ state instead of the logical $\ket{00}$ state.  Therefore, we need an additional step.  It is sufficient to perform CNOTs from qubits 1 and 4 to the ancilla, initialized in the state $\ket{0}$~\cite{Preskill}.  Then we measure the ancilla.  If there is no error, the ancilla should still be $\ket{0}$.  If we find the ancilla value $1$, we discard the run.

Note that it is important to use this particular circuit to create the cat state.  The single check is sufficient because of specific properties of the encoding circuit used, and another circuit to create the cat state might not work, even though it is equivalent in the absence of error.  In this case, we can note that a single gate error anywhere in the encoding circuit in the cat state can produce two bit flip errors in the cat state, but only at the cost of making the first and fourth qubits different, which shows up in the ancilla test.  It is also possible to get four bit flip errors, but that is equivalent to no error at all for this state.  Similarly, there is no possibility of two phase errors: two $Z$ errors brings us back to the correct cat state.  Thus, a single gate error in the encoding circuit either leaves a single-qubit error in the code (which will be detected in the final measurement step since it switches the parity of the outcome to odd), leaves no error at all, or is detected by the ancilla qubit.

The family of original circuits that can be encoded thus consists of preparation of $\ket{00}$, $\ket{0+}$, or $\ket{00} + \ket{11}$, followed by an arbitrary string of Pauli gates, SWAP with Hadamard on both qubits, and controlled-$Z$ gates.  Then the circuits are completed by measuring the two logical qubits in the standard basis.  Because we cannot do repeated error correction with this set-up, we should limit the length of these circuits to some reasonable size, e.g.~$10$ or $100$.  The error rate per gate needed to demonstrate fault tolerance decreases with the upper bound on the circuit length~\cite{Supplementary}, from a few percent for circuits of length $1$ to about $10^{-3}$ for circuits of length $100$.  The Supplementary Material~\cite{Supplementary} discusses how to add CNOT from the first logical qubit to the second one and repeated error detection using some additional connections between qubits, further expanding the repertoire of possible circuits.  

\section{Challenges and opportunities}

There are a variety of difficulties involved in verifying that a system meets the definition of fault tolerance proposed in this paper besides the obvious one of implementing the required quantum circuits with sufficient accuracy.  First, many of the circuits in the family have outputs which are not deterministic.  Possibilities include a uniformly random distribution on one or both of the two logical qubits.  Verifying that the output distribution is in fact uniform requires substantially more statistics than verifying a deterministic output, particularly since in a system with sufficiently high fidelity to have a chance at being fault tolerant, even the unencoded version of the circuit will have a relatively low error rate.  Moreover, in order to compute the error rate, one must know what the ideal circuit does.  If the original circuit only involves two logical qubits, as in this paper, that is not a problem, but is a serious issue if one attempts to scale up this sort of test to larger systems.

The biggest problem, however, is the requirement that the error rate should be lower for \emph{all} circuits in the family.  Even with a cutoff on the number of gates, the number of possible circuits is exponential in the cutoff.  The natural solution is to sample randomly circuits of varying sizes and to check the criterion on those circuits.  It is possible that certain non-random circuits behave worse than random ones (for instance, allowing coherent build-up of errors), so we should also test some representative set of non-random circuits too.  See the Supplementary Material~\cite{Supplementary} for a more specific suggestion as to what set of circuits to test.

Ideally, we would like to have a canonical set $S$ of circuits so that a system that has a lower logical error rate for every circuit in $S$ would be guaranteed to be fault-tolerant for all circuits in the full family.  Unfortunately, there does not appear to be any such set in general, since it is possible in principle to have non-Markovian error models which leave the error rate arbitrarily low for small circuits but become large after a sufficiently long time or will only reveal themselves in certain very specific situations.  A rigorous solution therefore requires some sort of assumption about the error model, but it is difficult to come up with an assumption which is certainly true for the experimental system under examination and yet is tractable enough to get any kind of result.
%

For the purposes of analyzing fault-tolerant protocols, we are not necessarily interested in learning the precise experimental error model.  Instead, we only want to know whether the errors in the system are of a nature that are handled well by the fault-tolerant protocol or poorly, and perhaps how poorly.  For instance, we expect independent depolarizing errors to be handled well by a fault-tolerant protocol, whereas two-qubit correlated errors can be corrected by a sufficiently large fault-tolerant protocol, but will presumably require additional resources.

The most important question is whether we can expect fault tolerance to work in a scaled-up version of the system. Darmawan et al.~\cite{DIP} found that none of the usual error metrics gives a good prediction of how well quantum error correction will perform on a channel.  Fault-tolerant experiments perhaps provide a way around this issue: It seems plausible that the best way to predict how well fault tolerance will work on a large system is to see how well it works on a small system.  To do so, it would be useful to find a way to quantify separately the amount of error in an experiment that is handled well by fault tolerance and the amount of error that is handled poorly.  One way to do this would be to try simulations of various specific error models and compare to the experiment, but it would be highly desirable to find better ways that don't depend on large simulations or strong assumptions about the nature of the errors.

In addition, experiments on fault tolerance offer an exciting opportunity to answer theoretical questions about the behavior of fault-tolerant protocols under different error models.  Classical simulation of large systems with non-Pauli error models is usually computationally intractable, but a fault tolerance experiment can provide the answer.  For instance, the behavior of coherent errors in a fault-tolerant protocol remains an open problem: Rigorous threshold proofs for coherent and non-Markovian errors~\cite{AGP} are based on bounding an operator norm such as the diamond norm.  This translates to a very stringent requirement on the fidelity of gates under coherent errors.  Conceptually, this occurs because coherent errors can add together amplitudes: If one gate has over-rotation $\theta$, corresponding to error probability $p \approx \theta^2$, then a sequence of $N$ gates could produce an over-rotation of $N \theta$ and an error probability of about $N^2 p$, whereas $N$ gates with incoherent errors of the same size would only have an error probability about $Np$.  However, in order to get this behavior, the errors need to add together in a completely coherent way.  It seems likely that there is limited ability for them to do so in the context of a fault-tolerant protocol, which involves many additional gates, such as CNOTs to measure the error syndrome.  Therefore, it seems likely that coherent errors are in fact no worse for fault tolerance than incoherent errors.  While this is a purely theoretical question, experiments would be very illuminating.

In short, the first experimental demonstrations of fault tolerance should be possible using quantum computers with current capabilities or just slightly beyond.  However, any demonstration of fault tolerance, to be fully convincing, will need to carefully study a variety of different encoded circuits and compare to unencoded circuits on the same hardware.  For larger-scale fault-tolerant experiments, it is unclear how to keep all needed resources within reason, but if this problem can be solved, there is an opportunity for fault-tolerant experiments to be not just \emph{demonstrations} of adequate control but to actually be \emph{experiments} teaching us something new about both the physical systems being studied and the theory of fault-tolerant quantum computation.

\paragraph*{Acknowledgements} I would like to thank many people who I have discussed this work with, including Hector Bombin, Ben Criger, Steve Flammia, Jay Gambetta, John Martinis, Oskar Painter, and Mark Tame.   Particular thanks goes to John Preskill for pointing out the possibility of repeated error detection described in the Supplementary Material.  Research at Perimeter Institute is supported by the Government of Canada through the Department of Innovation, Science and Economic Development Canada and by the Province of Ontario through the Ministry of Research, Innovation and Science.

\appendix

\section{Supplementary Material}

\subsection{Error rate calculation}

In order to determine the approximate error rate needed to demonstrate fault tolerance, I will use the approach of \cite{AGP} applied to the circuit of figure~\ref{fig:fourFT}.  However, in this case, there is only a single rectangle since we do not do repeated error correction.  The error rate needed will likely depend on the precise error model, but for simplicity, I will consider Pauli channels.

First, we count the total number of locations in the circuit.  Assuming that all qubits can be initialized with a $\ket{0}$ preparation just when they are needed, we have a total of $5$ $\ket{0}$ preparation locations, $5$ CNOT gate locations, $1$ Hadamard location, $2$ wait locations, and $5$ measurement locations, plus whatever gates are used in $U$.  Assuming $U$ consists of $T$ time steps, each with a transversal single-qubit gate, that gives a total of $A = 18 + 4T$ locations.  Using the simpler method of determining failure rates for rectangles, we find that, assuming local stochastic noise with physical error rate $p$, the logical error rate $p_L$ for the circuit (i.e., the probability that the rectangle is bad) is at most
\begin{equation}
p_L \leq \binom{A}{2}p^2 = (8T^2 + 70T + 153)p^2.
\end{equation}
However, we are post-selecting, and there is a fair chance that the run is rejected for showing an error.  This cannot happen if there is not at least one fault in the circuit, so the probability $p_S$ of surviving post-selection is at least
\begin{equation}
p_S \geq 1 - Ap = 1 - (18 + 4T) p.
\end{equation}

This circuit corresponds to a logical circuit with state preparation for two qubits, $T$ logical gates each on two qubits (perhaps a wait on one qubit and a gate on the other), and measurement of both logical qubits.  Therefore, the error rate $p_U$ for an unencoded circuit is $p_U \approx (4 + 2T)p$.  The unencoded circuit is never rejected, so to make a fair comparison of the logical and physical error rates, we should compare $p_L/p_S$ (the \emph{conditional probability} of logical error in the circuit) with $p_U$.

Consequently, the fault-tolerant circuit is better if
\begin{align}
\frac{(8T^2 + 70T + 153)p^2}{1-(18+4T)p} &< (4+2T)p \\
p &< \frac{4+2T}{16T^2 + 122T + 225}
\end{align}
For instance, for $T=1$, we get an improvement when $p < 6/363 \approx 1.6\%$.  However, the logical error rate increases slower with $T$ than the unencoded error rate, leading to a more stringent bound for larger $T$.  For instance, for $T=4$, we need $p < 12/969 \approx 1.2\%$.  For $T=10$, there is an improvement for $p < 0.79\%$, and for $T = 100$, we need $p < 0.12\%$.  This reflects the lack of repeated error correction or detection steps to keep the build-up of errors under control.  However, for small logical circuits, it appears that an error rate on the order of one percent or perhaps slightly less is sufficient.  These values can be improved further by counting malignant sets of errors, but the improvement is modest for these small circuits and doesn't change the overall picture. Restricting attention to depolarizing errors improves the bound on error rates by roughly a factor of $2$, since there is a chance that one of the two errors is of a type that will not influence the measurement outcome.

\subsection{Other codes and enhancements for the four-qubit code}

The $5$-qubit code~\cite{fivequbit1,fivequbit2} is the smallest possible code to correct one error, but fault-tolerant protocols for it~\cite{FTtheory} are not straightforward.  In particular, fault-tolerant state preparation and measurement are rather difficult and would require at least $11$ qubits in total even for a simple circuit with only state preparation and measurement.  The $7$-qubit code~\cite{CalderbankShor,Steane} has a simpler fault-tolerant protocol~\cite{ShorFT} but is larger to start with, and still requires $4$-qubit ancillas for error correction, plus an additional ancilla qubit to test the other ancillas, for a total of $12$ qubits.  Surface codes~\cite{Kitaev,DKLP} are very popular and are promising for fault tolerance in larger systems~\cite{recent-surface}, but the current proposals for experiments demonstrating fault tolerance with them~\cite{TSsmallsurface,WWLsurface} require $13$ or more qubits.  The $9$-qubit Bacon-Shor subsystem code~\cite{ninequbit,Bacon,Poulin} can get by with only one ancilla, so is actually one of the best candidates despite the relatively large code size, needing only $10$ qubits in total.  However, in order to implement the code using nearest-neighbor interactions in two dimensions, this would increase up to $13$ qubits.  Fault-tolerant protocols for the three-qubit phase-error-correcting code~\cite{AP} are quite complicated and in any case will not work by themselves unless all other sorts of errors are negligible.

The four-qubit code is a better option for an initial experiment, but it has only a limited set of gates that can be performed using the minimal resources assumed here, namely a ring of $5$ qubits.  However, with some additional connections between qubits, $5$ physical qubits can also allow a logical CNOT gate and repeated fault-tolerant error detection.

In particular, swapping physical qubits $1$ and $2$ does a CNOT from the first logical qubit to the second logical qubit.  The physical SWAP gate would not be fault-tolerant, since an error during it could introduce errors to two qubits, which might not be detected.  If we allow a conceptual SWAP (just relabeling the qubits), then we can do the CNOT.  Alternatively, if we have an additional connection between qubits $2$ and $A$, then a small circuit with 3 physical SWAP gates can implement the logical SWAP without risk of a two-qubit error on the two code qubits.  (However, in this case, we must have already used the ancilla qubit for its main purpose in state preparation and measured it.)

Another component of fault tolerance missing in this protocol is the ability to correct (or in this case, detect) errors and then continue with gates.  Using a slightly different arrangement of qubits, with $A$ located at the center of a square of the four qubits used in the code, it is also possible to add in-place error detection.  The code is slightly different.  It is now the four-qubit Bacon-Shor subsystem code and only encodes one logical qubit~\cite{Bacon,Poulin}.  The ancilla is still used for encoding in the same way as before, but it can also be used to measure the gauge operators, which can be put together to deduce the  error syndrome.  This geometry may be more difficult to implement, but allows much longer circuits with a relatively low logical error rate.  However, since the code only detects errors, the cost of doing long circuits is that the odds of successfully completing a run without detecting an error become much smaller.  Of course, even with repeated error detection and post-selection, in a long enough circuit the error will eventually overcome the code when two errors happen to occur in quick succession before error detection can identify them.

\subsection{Circuit subfamilies for demonstrating fault tolerance}

Suppose we have a circuit family consisting of circuits on $n$ qubits beginning with a standard state preparation (e.g., of the $\ket{0}$ state), ending with measurements of all qubits (perhaps in the standard basis), and with intermediate gates drawn from some set in all possible combinations.  If we have a total of $g$ different gates that can be performed at each time step (counting separately the same gate applied to different qubits and parallel implementation of different gate sets), then there are a total of $g^T$ circuits with depth $T+2$ (counting state preparation and measurement as one layer of operations each).  For instance, for the four-qubit code with the logical $\ket{00}$ preparation, we have a total of $g = 18$: $16$ logical two-qubit Paulis, the Hadamard/SWAP gate, and the controlled-$Z$ gate.  Even if we cut off the size of possible circuits at $T=10$, that still leaves over $10^{12}$ different possible circuits in the family.  Thus, testing all of the circuits in the family is prohibitively expensive.

Instead, we wish to find a small subfamily to test.  We want the subfamily to be sufficiently representative of the full family so that if fault tolerance improves the error rate for every circuit in the subfamily, then we will be convinced that it will do so for every circuit in the full family.  The subfamily should include both short and long circuits to test different ratios of gates to state preparation and measurement.  It should include both random circuits and non-random circuits: non-random circuits might be worse because of the build-up of coherent errors, but they could also be better because errors might be more likely to cancel out.

Therefore, I suggest the following subfamily $S$, based on constant parameters $T$ (the cutoff in circuit length for the full family), $r$ (the number of different circuit types to test of various kinds), and $p$ (maximum periodicity to test).
\begin{enumerate}
\item For each value from $t=0$ to $T$, choose $r$ random circuits with $t$ time steps and test those.  (For small $t$, $r$ might be greater than the number of possible circuits, so it is sufficient to check all possible circuits instead.)
\item For each period from $q=1$ to $p$, choose $r$ random circuits $C$ with $q$ time steps.  Test the circuits consisting of repetitions of $C$, starting with $1$ repetition and going up to $\lfloor T/q \rfloor$ repetitions.  (Again, for small periods $q$, $r$ may be greater than the number of possible circuits, so test all possibilities for $C$.)
\end{enumerate}
The total number of circuits in this subfamily is at most $r(T+1) + rT (\ln p + 1)$.  This is a much more reasonable size of subfamily to test.  If we want to allow multiple different initial preparations (e.g., $\ket{00}$, $\ket{0+}$, $\ket{00} + \ket{11}$ in the case of the four-qubit code), choose a subfamily for each initial state.  There seems to be no particular reason to insist either that subfamilies for different initial states be the same or that they be different; either will work.

For each circuit $C$ in the subfamily $S$, the experiment should implement both unencoded versions of $C$ and encoded fault-tolerant versions of $C$ and do both enough times to have good statistics on the outcome distribution. Suppose $\{p_i\}$ is the outcome distribution of the ideal circuit $C$ (computed on a classical computer), $\{q_i\}$ the experimentally measured outcome distribution of the unencoded version of $C$, and $\{r_i\}$ the experimentally measured decoded logical outcome distribution of the fault-tolerant version of $C$.  The measured unencoded error rate of $C$ is
\begin{equation}
P_u (C) = \frac{1}{2} \sum_i |p_i - q_i|,
\end{equation}
and the encoded error rate of $C$ is
\begin{equation}
P_e (C) = \frac{1}{2} \sum_i |p_i - r_i|.
\end{equation}
The experiment has demonstrated fault tolerance if $P_e(C) < P_u(C)$ (with sufficient confidence given statistical errors) for all $C \in S$.  

Note that there should be few systematic sources of error here: The effect of most experimental errors is considered part of the experiment, part of what is being measured.  Provided the unencoded version of each circuit is implemented in the same hardware as the encoded version, this is likely a fair comparison.  One possible source of systematic error would be a drift in calibration over the course of the full experiment.  This is not inherently a problem: A fault-tolerant system should be able to handle a range of error rates and still improve the error rate within a range of calibration errors.  However, since the analysis involves comparing error rates from different runs, it is possible that calibration drift could lead to some misleading results.  To minimize the effect of this, the unencoded version and the encoded version should be done close together in time so that they are comparing similar error rates.  In addition, it is wise to randomize the order in which circuits $C \in S$ are performed; otherwise, one might worry that drift will affect long circuits (done later) differently than short ones (done earlier), for instance.

\end{document}